\documentclass{article}
\usepackage{spconf,amsmath,graphicx,amsfonts,caption,subcaption}
\usepackage{graphicx}
\usepackage{caption}

\title{Large Deformation Diffeomorphic Metric Mapping and Fast-Multipole Boundary Element Method provide new insights for Binaural Acoustics}
\name{Reza Zolfaghari, Nicolas Epain, Craig T. Jin, Joan Glaun\`es, Anthony Tew
\thanks{This work was supported by the UK Engineering and Physical Sciences Research Council (grant GR/T28140/01) and the Australian Research Council's Discovery Projects funding scheme (project number DP110102920).}
\thanks{R. Zolfaghari, N. Epain and C.T. Jin are with CARLab, School of Electrical and Information Engineering, The University of Sydney, Sydney, Australia. email: craig.jin@sydney.edu.au.}
\thanks{A. Tew is  with the Department of Electronics, The University of York, Heslington, York, UK. email: tony.tew@york.ac.uk.}
\thanks{J.A. Glaun\`es is with the MAP5, Universit\'e Paris 5-Ren\'e Descartes, 75006 Paris, France. email: alexis.glaunes@mi.parisdescartes.fr.}
\address{}}

\begin{document}
%
\maketitle
\begin{abstract}
This paper describes how Large Deformation Diffeomorphic Metric Mapping (LDDMM) can be coupled with a Fast Multipole (FM) Boundary Element Method (BEM) to investigate the relationship between morphological changes in the head, torso, and outer ears and their acoustic filtering (described by Head Related Transfer Functions, HRTFs).  The LDDMM technique provides the ability to study and implement morphological changes in ear, head and torso shapes. The FM-BEM technique provides numerical simulations of the acoustic properties of an individual's head, torso, and outer ears.  This paper describes the first application of LDDMM to the study of the relationship between a listener's morphology and a listener's HRTFs. To demonstrate some of the new capabilities provided by the coupling of these powerful tools, we examine the classical question of what it means to ``listen through another individual's outer ears.''  This work utilizes the data provided by the Sydney York Morphological and Acoustic Recordings of Ears (SYMARE) database.
\end{abstract}
\section{Introduction}
\label{sec:intro}
Morphoacoustic signal processing is a relatively new term in binaural acoustics~\cite{Tew2012}. It refers to signal processing in which the relationship between a listener's morphology (the shape of the torso, head, and ears) and the listener's individualized acoustic filtering properties plays a prominent role. The acoustic filters required for synthesizing high-fidelity 3D audio for an individual listener are referred to as head-related impulse response (HRIR) filters. These filters describe the acoustic filtering properties of the torso, head, and outer ears and how they transform the sound waves that arrive from some location in space, interact with the ear, and ultimately reach the tympanic membrane. This acoustic filtering imparts a signature to the incoming sound that the human auditory system perceptually decodes as spatial information.  The Fourier transform of the HRIR filters are referred to as head-related transfer functions (HRTFs). 

The study of the relationship between morphological shape and acoustic filtering has been on-going for over a decade now. For example, \cite{Jin2000} applied regression and principle component analysis to establish a linear relation between morphological and acoustic data. In another study, \cite{Zotkin2003} proposed a ``best fit'' HRTF based on a similarity score between individual anthropometric measurements and data available in the CIPIC database. More recently \cite{Tew2012} has characterized the effect of small perturbations on the magnitude of the notches and peaks appearing in the frequency spectrum of a single HRTF. Similar work conducted by \cite{Mokhtari2011} investigates the effect of the ear morphology upon the magnitude of the N1 notch in the median plane.

We now briefly introduce the tools that are applied in this study, beginning with the FM-BEM. Traditionally, individual HRTFs are acoustically measured in an anechoic chamber that incorporates a robotic arm to move a sound source in space. Microphones are placed in the ears of the listener and impulse responses are recorded for a number of directions in space. This clearly requires specialized equipment and is time consuming for the listener. With the increasing computational power of personal computers and also the improvement in numerical methods for boundary element method simulations, such as the fast multipole method \cite{Gumerov2010}, a listener's HRTFs can now be numerically derived using acoustic simulations conducted using high-resolution surface meshes of the listener.  Considerable research has been done to establish the validity of numerical simulations of HRTFs using the BEM~\cite{Gumerov2010,Jin2013,Kahana2007}. These research show that high resolution surface meshes of the torso, head and ears are sufficient to provide accurate numerical simulation of HRTFs that provide a reasonable match to the acoustically-measured HRTFs.

\begin{figure*}
\centering
	\centerline{\includegraphics[width=\textwidth]{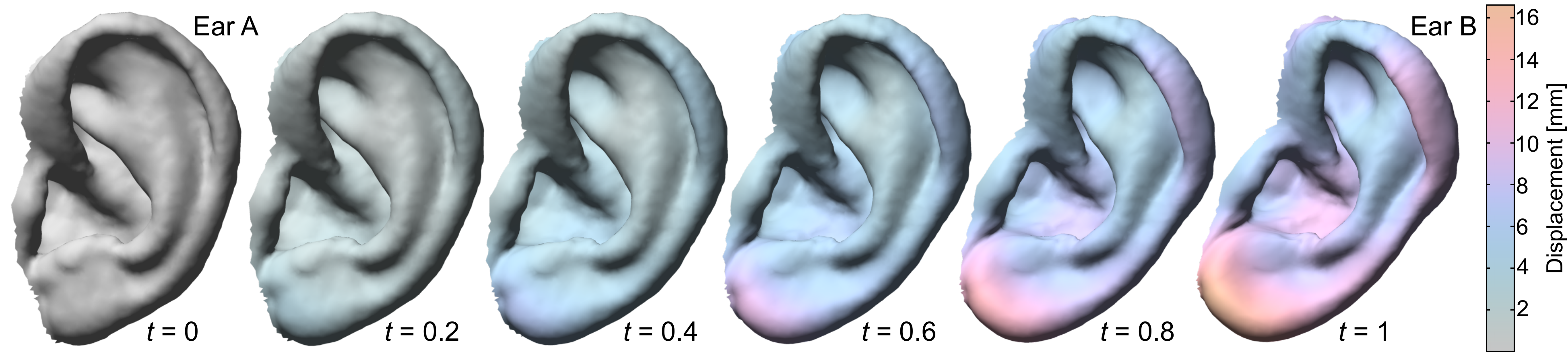}}
	\captionof{figure}{The result of the flow of diffeomorphisms for several time steps is shown for the matching of Ear~A to Ear~B. The color indicates cumulative displacement. A contant luminance colormap has been used for best results, so please examine the figure online.}
	\label{fig:flow}
\end{figure*}

We now introduce a shape analysis tool referred to as Large Deformation Diffeomorphic Metric Mapping (LDDMM)~\cite{Joshi2000,Beg2005} that will be used to manipulate and study ear shapes. LDDMM models the mapping of one surface, $C$, to another surface, $S$, as a dynamic flow of diffeomorphims of the ambient space, $\mathbb{R}^3$, in which the surfaces are embedded. The computations for LDDMM are not performed directly in the space of diffeomorphisms; instead, the method works with time dependent vector fields, $\mathbf{v}(t): \mathbb{R}^3\rightarrow \mathbb{R}^3$ for t $\in [0,1]$, which model the infinitesimal efforts of the flow. The flow of diffeomorphisms, $\phi^\mathbf{v}(t,X)$, operating on a subset $X \subset \mathbb{R}^3$ is defined via the partial differential equation:
\begin{equation}
\frac{\partial \phi^\mathbf{v}(t,\mathbf{X})}{\partial t} = \mathbf{v}(t)\circ\phi^\mathbf{v}(t,\mathbf{X}) \ \text{,}
\end{equation}
where $\circ$ denotes the composition of functions. Note that the superscript $\mathbf{v}$ on $\phi^\mathbf{v}(t,\mathbf{X})$ has no significance except to denote that the flow of diffeomorphisms is defined for a particular set of time dependent vector fields $\mathbf{v}(t)$.

The diffeomorphism at time $t=0$, is the identity diffeomorphism: $\phi^\mathbf{v}(0,C)=C$.  The result of the flow of diffeomorphisms at time $t=1$ maps $C$ to $S$: we write this as $\phi^\mathbf{v}(t, C)|_{t:[0\rightarrow1]} = S$. The time dependent vector fields, $\mathbf{v}(t)$, belong to a Hilbert space of regular vector fields that has a kernel, $k_V$, and a norm $\|\cdot\|_V$ that models the infinitesimal cost of the flow. In the LDDMM framework, we solve an inexact matching problem which minimizes the cost function, $J_{C,S}$, defined as:
\begin{align}
& J_{C,S}\left(\mathbf{v}(t)_{t\in[0,1]}\right) = \gamma\int_0^1\|\mathbf{v}(t)\|_V^2dt \nonumber \\ & \hspace{1.2in} + E\left(\phi^\mathbf{v}(t, C)|_{t:[0\rightarrow1]},S\right) \ \text{,}
\end{align}
where $E$ is a norm-squared cost measuring the degree of matching between $\phi^\mathbf{v}(t, C)|_{t:[0\rightarrow1]}$ and $S$. In this work, we use the Hilbert space of currents~\cite{Joan2008,Vaillant2005,Vaillant2007} to compute $E$ because it is easier and more natural than using landmarks. 

In order to make the concept of the flow of diffeomorphisms explicit, consider Fig.~\ref{fig:flow}. In this figure, we transform the surface mesh for Ear~A to that for Ear~B. We show the flow of diffeomorphisms $\phi^\mathbf{v}(t, \mathrm{Ear \ A})|_{t:[0\rightarrow t_i]}$  for times $t_i = \{0, 0.2, 0.4, 0.6, 0.8, 1.0\}$.

In order to apply the tools of FM-BEM and LDDMM, we use the SYMARE database~\cite{Jin2013} which provides high-resolution surface meshes of the head and torso both with and without ears. The database also provides high-resolution surface meshes of the ears alone. The basic idea behind this paper is to manipulate shapes using LDDMM and to examine the resulting acoustics using FM-BEM. The problem that we consider is the classical question of: ``what does it mean to listen through another individual's outer ears.'' In examining this question, we  demonstrate and validate our approach to combining the tools of LDDMM and FM-BEM. We also reveal an interesting finding regarding the influence of the head and torso on HRTFs. In Section 2 we describe the methods; Section 3 describes the results and Section 4 concludes.

\section{Method}
\label{sec:method}
\begin{figure}[t]
  \centering
  \centerline{\includegraphics[width=.3\textwidth]{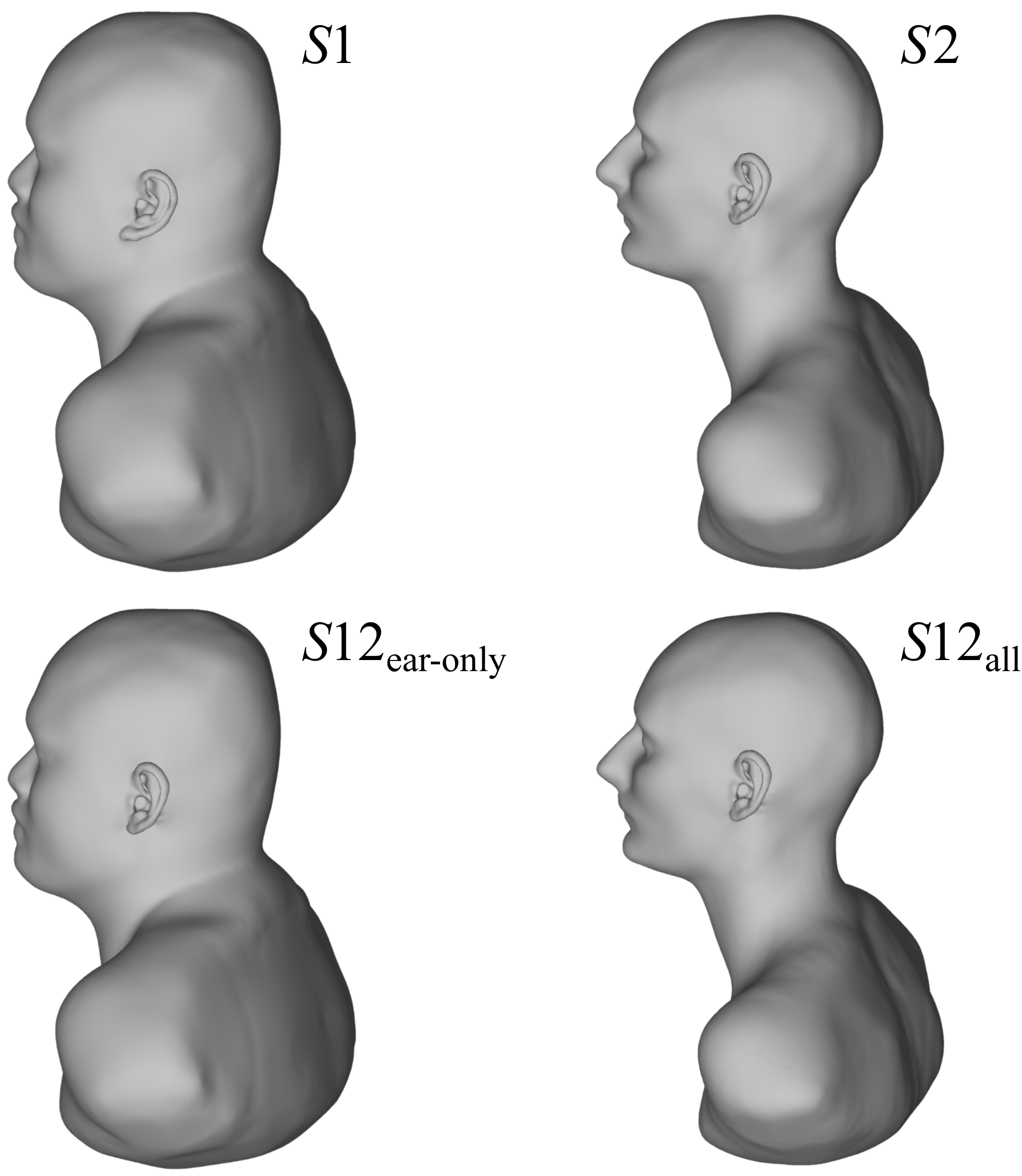}}
 \caption{Surface meshes are shown for $S1$ and $S2$ (top row) and for $S12_{\mathrm{ear-only}}$ and $S12_{\mathrm{all}}$ (bottom row).}
\label{fig:HTs}
\end{figure}

This study starts with high-resolution surface meshes (approximately 800 K triangular elements) of the torso, head, and ears for two subjects in the SYMARE database, referred to as $S1$ and $S2$ (top row of Fig.~\ref{fig:HTs}). The surface meshes are triangulated meshes composed of a collection of vertices, $x_n,1\leq i \leq N$, such that $x_n \in R^3$ and a set of connectivity information for these points, $f_m,1\leq m \leq M$, which constitute the triangular faces of the mesh. Note that for simplicity, we use $S1$ and $S2$ to designate both the subjects and their corresponding surface meshes. Using the tools of LDDMM, we apply two shape transformations to the surface mesh for $S1$. In one transformation, we transform the left ear of $S1$ such that it is similar to the left ear of $S2$. The resulting surface mesh is referred to as $S12_{\mathrm{ear-only}}$. In a second transformation, we transform the torso, head, and left ear (but not the right ear) of $S1$ to be similar to the torso, head, and left ear of $S2$. The resulting surface mesh is referred to as $S12_{\mathrm{all}}$. The two shape transformations described above are shown in the bottom row of Fig.~\ref{fig:HTs}. We also perform similar shape transformations to $S2$, but in the reverse direction from $S2$ to $S1$, to obtain $S21_{\mathrm{ear-only}}$ and $S21_{\mathrm{all}}$. All together, we have six different surface meshes consisting of the two original surface meshes and the four transformed surface meshes.

In order to describe how the shape transformations are implemented using LDDMM, we define three mathematical operations -- \emph{translate}: $\mathcal{T}$; \emph{match}: $\mathcal{M}$; and \emph{flow}: $\mathcal{F}$. The mathematical operation $\mathcal{T}(S1,S2)$  applies a translation to the surface mesh $S1$ such that it best matches surface mesh $S2$. The mathematical operation $\mathcal{M}(S1,S2)$ finds the momentum vectors, $\boldsymbol{\alpha}_n(t)$, that minimize the cost function $J_{S1,S2}\left(\mathbf{v}(t)_{t\in[0,1]}\right)$. The momentum vectors describe the change to a point,  $\mathbf{x}(t)$, in the ambient space,  at a given time step via the following differential equation:
\begin{equation}\label{equ:flow}
\frac{d\mathbf{x}(t)}{dt}=\sum_{n=1}^N k_V(\mathbf{x}_n(t),\mathbf{x})\boldsymbol{\alpha}_n(t) \ \text{,} 
\end{equation}
where $k_V(\mathbf{x},\mathbf{y})$ is the Cauchy kernel defined by:
\begin{equation}
k_V(\mathbf{x},\mathbf{y}) = \frac{1}{1+\frac{\|\mathbf{x}-\mathbf{y}\|^2}{\sigma_V^2}} \ \ \text{.}
\end{equation}
The parameter $\sigma_V$ determines the range of influence of the momentum vectors $\boldsymbol{\alpha}_n(t)$. The mathematical operation $\mathcal{F}\left(S1,\left\{\boldsymbol{\alpha}\right\}\right)$ applies the flow of diffeomorphisms defined by a set of momentum vectors, $\left\{\boldsymbol{\alpha}\right\}$, to the surface mesh $S1$.  For simplicity in this notation, we have not made explicit the time dependence of the momentum vectors, $\left\{\boldsymbol{\alpha}\right\}$.

\begin{figure}[htb]
  \centering
  \centerline{\includegraphics[width=8.5cm]{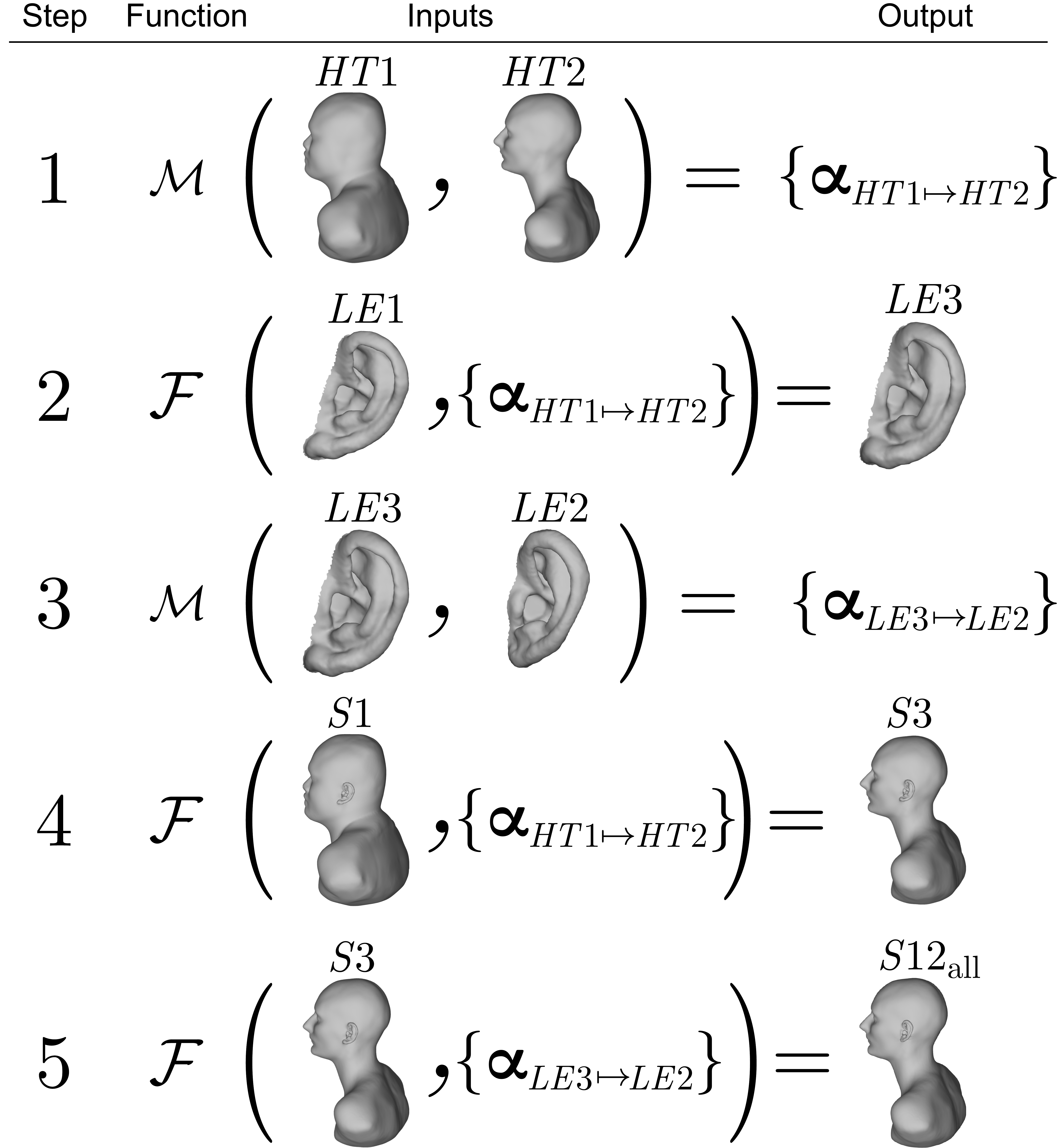}}
\caption{The steps in the transformation of $S1$ to $S12_{\mathrm{all}}$ are shown.}
\label{fig:Table1}
\end{figure}

\begin{figure}[!!!!h]
  \centering
  \centerline{\includegraphics[width=8.5cm]{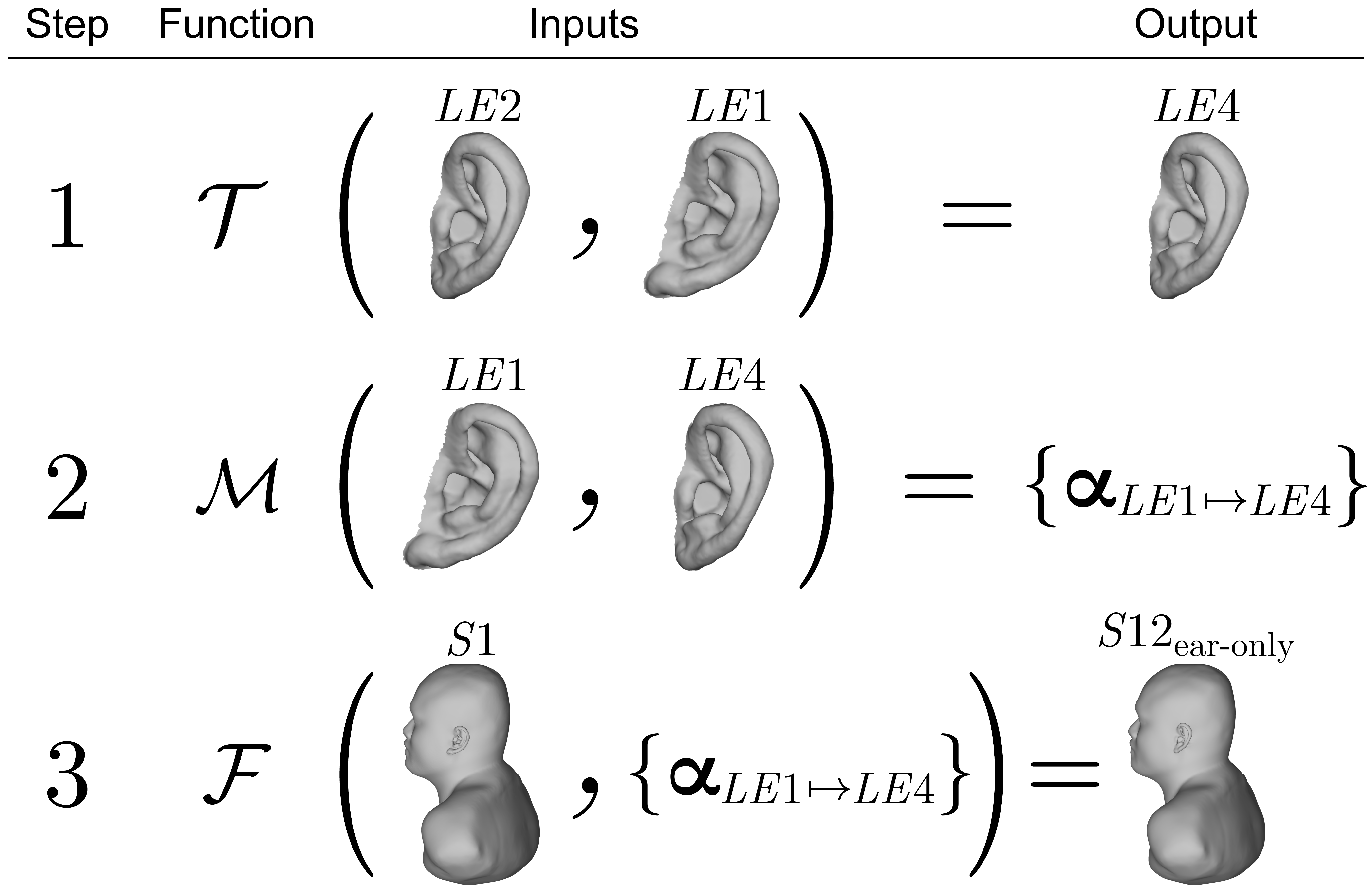}}
\caption{The steps in the transformation of $S1$ to $S12_{\mathrm{ear-only}}$ are shown.}
\label{fig:Table2}
\end{figure}

\begin{figure*}[t]
  \centering
  \centerline{\includegraphics[width=\textwidth]{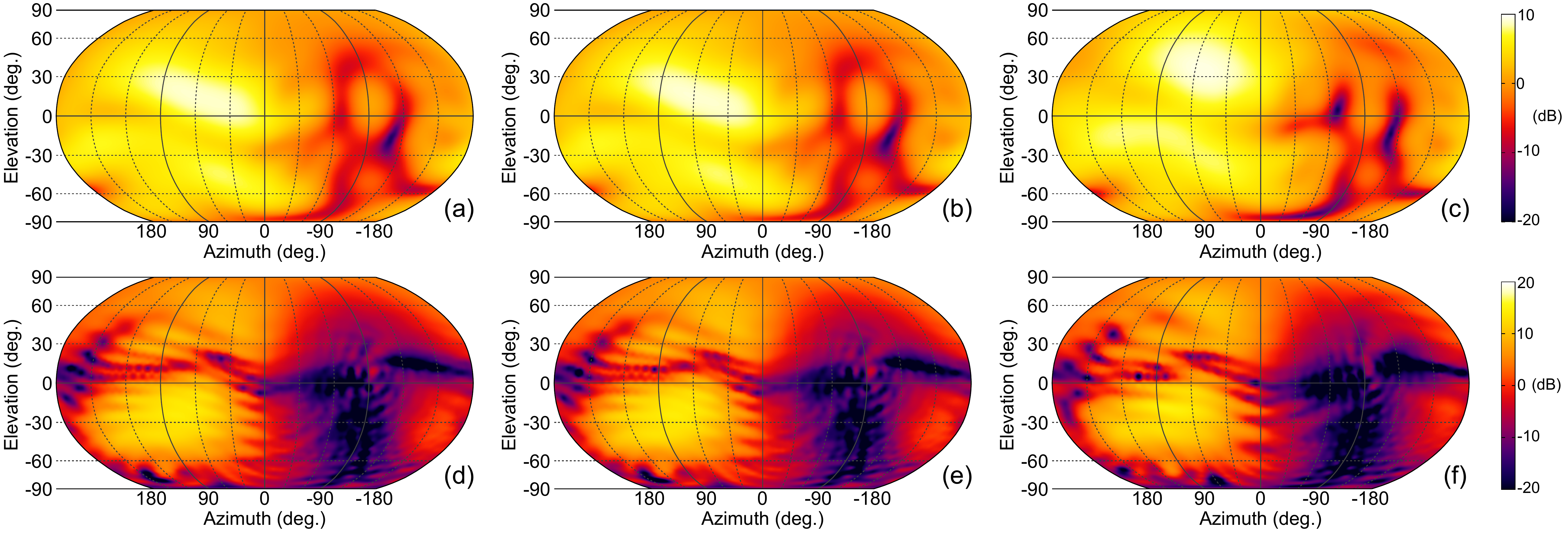}}
\caption{SFRS plots for $S2$ (a,d); $S12_{\mathrm{all}}$ (b,e); and $S12_{\mathrm{ear-only}}$ (c,f) are shown at two frequencies, 2~kHz (top row) and 10~kHz (bottom row).}
\label{fig:SFRS}
\end{figure*}

With the definitions of the three mathematical operations ($\mathcal{T}$, $\mathcal{M}$, and $\mathcal{F}$)
in hand, it is straightforward to describe the methods to determine $S12_{\mathrm{ear-only}}$ and $S12_{\mathrm{all}}$ from $S1$. Fig.~\ref{fig:Table1} shows and describes the five steps required to synthesize  $S12_{\mathrm{ear-only}}$ from $S1$. Fig.~\ref{fig:Table2} shows and describes the three steps required to synthesize $S12_{\mathrm{all}}$ from $S1$. In order to follow the procedures listed in Figs.~\ref{fig:Table1}~and~\ref{fig:Table2}, it is important to understand that the SYMARE database provides high-resolution surface meshes of the torso and head without ears. So, for example, in Fig.~\ref{fig:Table1} we can learn the momentum vectors required to match $HT1$ (the torso and head surface mesh of $S1$ without ears) to $HT2$ (the torso and head surface mesh of $S2$ without ears). We can then apply this flow of diffeomorphisms to $LE1$ (the left ear of $S1$) to obtain the intermediate left ear $LE3$. We can then learn the momentum vectors required to match $LE3$ to $LE2$ (the left ear of $S2$). We can then apply both flows of diffeomorphisms sequentially to $S1$ to obtain $S12_{\mathrm{all}}$. Synthesizing $S12_{\mathrm{all}}$ is a control condition in the sense that the HRTFs for $S12_{\mathrm{all}}$ should be identical to that for $S2$. In Fig.~\ref{fig:Table2} we show the steps to obtain $S12_{\mathrm{ear-only}}$. Because the head diameter of $S1$ and $S2$ are not identical, we translate $LE2$ to match $LE1$. We then learn the momentum vectors required to match $LE1$ to the translated version of $LE2$ and apply the flow of diffeomorphisms to $S1$ to obtain $S12_{\mathrm{ear-only}}$. 


In order to obtain the HRIRs corresponding to the six surface meshes ($S1$, $S2$, $S12_{\mathrm{ear-only}}$, $S12_{\mathrm{all}}$, $S21_{\mathrm{ear-only}}$, $S21_{\mathrm{all}}$) we apply FM-BEM simulations. In this work we used the Coustyx software by Ansol. The simulations were performed by the FM-BEM solver using the Burton-Miller Boundary Integral Equation (BIE) method.  Using the acoustic reciprocity principle, a single simulation is used to determine all of the HRIRs in one go by placing a source on a surface mesh element that forms part of the blocked ear canal and then setting a uniform normal velocity boundary condition on this surface element. A post-processing step was used to refine the meshes prior to the FM-BEM simulation using the open-source software ACVD. The criteria the meshes need to meet during the mesh refinement are described in~\cite{Jin2013}.

\section{Results}
\label{sec:results}
\begin{figure}[]
        \centering
        \centerline{\includegraphics[width=0.5 \textwidth]{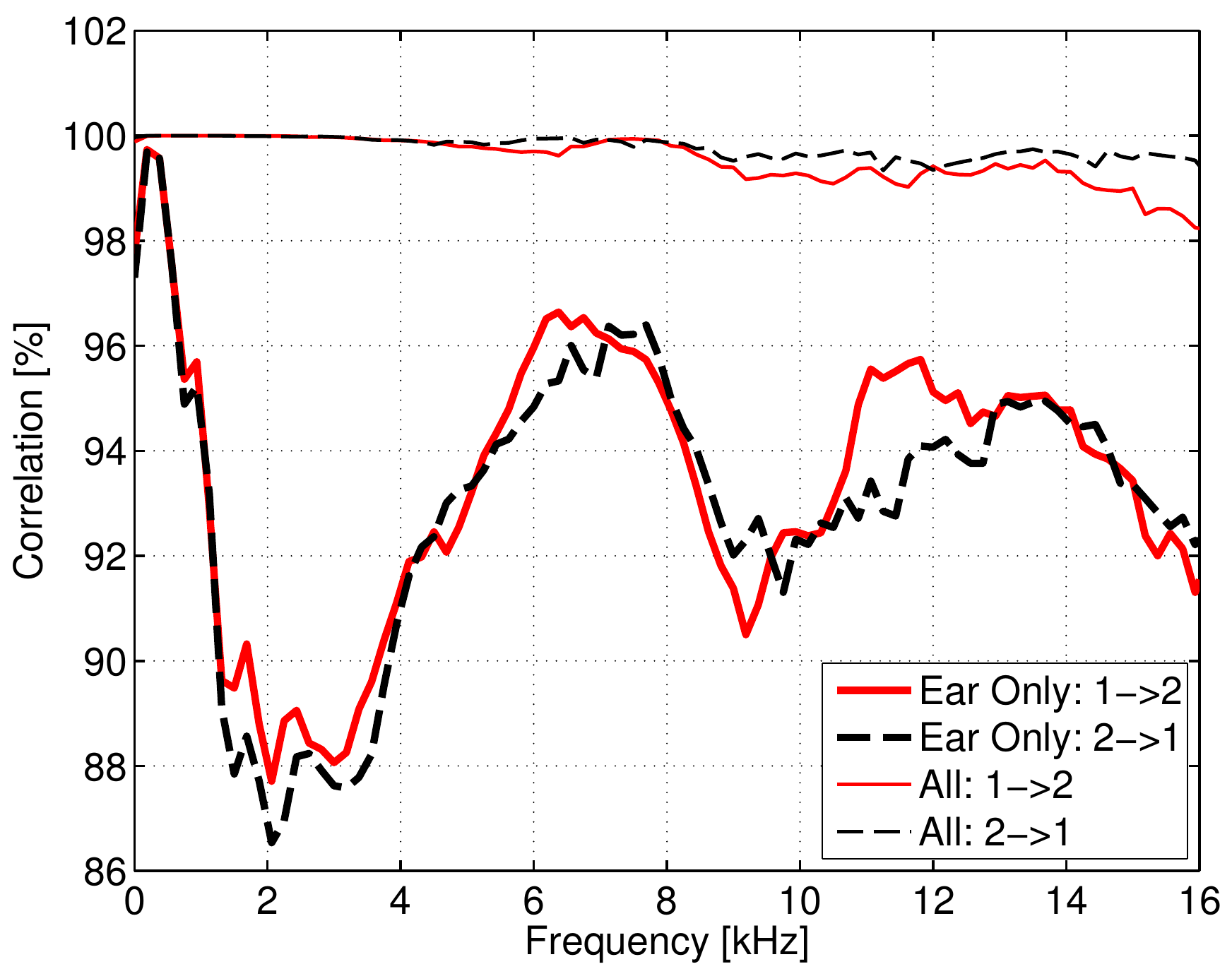}}
        \caption{The spatial correlation between various SFRS's are shown as a function of frequency. The spatial correlation is shown for the following pairing: ($S2$ and $S12_{\mathrm{ear-only}}$) and ($S2$ and $S12_{\mathrm{all}}$) -- solid line;  ($S1$ and $S21_{\mathrm{ear-only}}$) and ($S1$ and $S21_{\mathrm{all}}$) -- dotted line.}
\label{fig:corrCoef}
\end{figure}

We now compare the FM-BEM simulated HRIR data for $S2$, $S12_{\mathrm{ear-only}}$, $S12_{\mathrm{all}}$. In order to make these comparisons we plot the spatial frequency response surfaces (SFRS) corresponding to the HRTF data. An SFRS plot (see~\cite{Cheng1999} for details) shows the magnitude gain of the HRTF for a single frequency as a function of direction in space. Fig.\ref{fig:SFRS} shows the SFRS plots for $S2$, $S12_{\mathrm{ear-only}}$, and $S12_{\mathrm{all}}$ at 2~kHz and 10~kHz. The SFRS plots for $S2$ and $S12_{\mathrm{all}}$ are pretty much identical, while the SFRS plot for $S12_{\mathrm{ear-only}}$ shows differences that can be attributed to the different torso and head. The spatial correlation between the SFRS's for $S2$ and $S12_{\mathrm{all}}$ and between the SFRS's for $S2$ and $S12_{\mathrm{ear-only}}$ was calculated as a function of frequency and are shown in Fig~\ref{fig:corrCoef}. Morphological differences in the torso and head causes the spatial correlation to dip around 2~kHz and somewhat surprisingly around 9~kHz. 

\section{Conclusions}
\label{sec:conc}
In this paper we have demonstrated the first application and combination of the tools of LDDMM and FM-BEM to gain insights into binaural acoustics. We have shown how the tools can be applied to examine the classical question of ``listening through another individual's outer ear.'' LDDMM provides a powerful and flexible tool to study, characterize and manipulate ear shapes. In future work, we will statistically characterize the distribution of torso, head and ear shapes and their relationship to changes in binaural acoustics.

\bibliographystyle{IEEEbib}
\bibliography{icassp2014}

\end{document}